\documentclass[12pt]{article}
\usepackage{hhline}
\usepackage{latexsym,graphicx}
\usepackage{subfigure}
\usepackage{amssymb}
\usepackage{amsmath}
\usepackage{amscd}
\usepackage{amsthm}
\usepackage{float}
\usepackage{xcolor}
\usepackage[left=2cm,top=2.5cm,right=2.5cm,bottom=1.5cm]{geometry}   
 
\begin{document}
\begin{center}
\large{\bf{Logamediate inflation on the Swiss-cheese brane with varying cosmological constant}} \\
\vspace{10mm}
\normalsize{Nasr Ahmed$^{1,2}$, Anirudh Pradhan$^3$}\\
\vspace{5mm}
\small{\footnotesize $^1$ Mathematics Department, Faculty of Science, Taibah University, Saudi Arabia.} \\
\small{\footnotesize $^2$ Astronomy Department, National Research Institute of Astronomy and Geophysics, Helwan, Cairo, 
	Egypt}\\
\small{\footnotesize $^3$ Centre for Cosmology, Astrophysics and Space Science (CCASS), GLA University, Mathura-281 406, Uttar Pradesh, India}\\

\vspace{2mm}

$^{1,2}$E-mail: nasr.ahmed@nriag.sci.eg \\
$^3$E-mail: pradhan.anirudh@gmail.com \\

\end{center}  
\date{}
\begin{abstract}
The existence of Schwarzchild black holes in the structure of Swiss-cheese brane-world led to the conclusion that this specific brane-world scenario is more realistic than the FLRW branes. In this paper, we show that a Logamediate inflation on the Swiss-cheese brane with time-dependent cosmological constant $\Lambda(H)$ leads to a positive kinetic term and a negative potential with AdS minimum. The cosmic pressure $p$ is always positive but the energy density $\rho$ starts to get negative after a finite time. However, there is a time interval where they both are positive. Although this behavior of $\rho$ can be considered as a drawback of Swiss-cheese brane where positive energy dominates the present universe, it has been suggested that the presence of some source of negative energy could have played a significant role in early cosmic expansion. The model suffers from the eternal inflation problem which appears from the evolution of the first slow-roll parameter $\epsilon$. Due to the existence of $\rho^2$ term, we have tested the new nonlinear energy conditions. The slow-roll parameters have been investigated and compared to Planck 15 results.  
\end{abstract}

\section{Introduction}
The current accelerating expansion of the universe has been indicated by numerous observations \cite{11,13,14}. A mysterious form of energy with negative pressure, dubbed as dark energy (DE), has been assumed as a possible explanation where the negative pressure acts as a repulsive gravity. Some dynamical scalar fields models of dark energy have been proposed including quintessence \cite{quint}, Chaplygin gas \cite{chap}, phantom energy \cite{phant}, k-essence \cite{ess} and tachyon \cite{tak}. The accelerated expansion has also been studied by modifying the geometrical part of the general relativity action \cite{moddd}. Examples of this direction include $f(R)$ gravity \cite{39} where $R$ is the Ricci scalar, Gauss-Bonnet gravity \cite{noj8}, $f(T)$ gravity \cite{torsion} where $T$ is the torsion scalar, and $f(R,T)$ gravity \cite{1}, where $T$ is the trace of the energy momentum tensor. A detailed review of dark energy in modified gravity has been given in \cite{add1}. \par
The concept of inflation was initially introduced to elucidate certain cosmological issues, including the horizon problem, homogeneity problem, and monopole problem \cite{A4,A5,A6}. Due to the issue of internal manifold compactification stability, inflationary models were a significant challenge to construct in string theory. Consequently, a variety of models have been developed using distinct methodologies, including ``brane inflation." Brane-worlds is an extra-dimensional proposal for modified gravity in which the universe is a $3+1$-dimensional hyper-surface embedded in a higher dimensional space-time, $3+1+n$, called the bulk \cite{n2,n22}. The concept of a domain wall universe was proposed for the first time in \cite{domain}. Other proposals for brane-worlds scenario have been presented such as the universal extra dimension model \cite{yours}, the DGP model \cite{brane6ii}, GRS model \cite{brane6i} and the thick brane model \cite{brane6iii}. The phantom behavior free from the big rip singularity and the geometrical self acceleration are features of brane-world cosmology \cite{brane6iiii4}.\par An important feature of brane-world cosmology is the existence of a quadratic energy density $\rho^2$ term which can be of great relevance for building inflationary models on the brane. For anisotropic space-times, this $\rho^2$ term leads to a basic change in the nature of the initial singularty (big bang) \cite{ukra}. At low energy, brane-world models behave like scalar tensor theories \cite{derahm}. In the early universe, the low energy condition can't be applied and the 5-dimensional description becomes necessary. That makes analytical solutions impossible and approximation methods will be required  \cite{derahm}. \par

For the sake of completeness of this discussion, we have to mention that the interest in inflationary stages motivated by String Theory, albeit fundamental, has lost partially its strength in recent years. According to the Planck mission, string potentials are disfavored than Starobinsky's one and/or than non-minimally coupled inflationary potential. \par
The main motivation behind the current work is to investigate the effect of the quadratic energy term in the cosmological equations of the Swiss-cheese brane-world on the logamediate inflation scenario. Then, we study the evolution of the cosmic pressure, energy density, kinetic term and potential. The main purpose behind Swiss-cheese cosmology is to construct locally inhomogeneous space-times which appear globally isotropic and satisfy Einstein equations. The current work represents a fair study to see what cosmic behavior we can get by implementing the logamediate scale factor in the Swiss-cheese cosmological equations. Another motivation is that the logamediate inflation has been investigated by Barrow \& Nunes (2007) in the context of the standard inflationary scenario based on General relativity. Their results shows that logamediate inflation, within the standard inflationary framework, is ruled out by the Planck 2015 observations which opens the door for examining logamediate inflation in different modified gravity theories. \par
The paper is organized as follows: In section 2 we give a quick review for the Swiss-cheese brane-world and its cosmological equations. In section 3 we use the logamediate inflation scale factor with the cosmological equations to get the expressions for pressure and energy density. We then calculate the kinetic energy and potential. Section 4 is dedicated to investigate the physical acceptability of the model through energy conditions and causality. In Section 5 we calculate the slow-roll parameters, and in section 6 we study the cosmography. The conclusion and future perspectives are included in section 7.
\section{Swiss-cheese brane-world cosmology}
The Einstein equations on the brane have been derived in \cite{eins} and can be written as
\begin{equation}
G_{ab}=-\Lambda g_{ab}+\kappa^2T_{ab}+\tilde{\kappa}^4S_{ab}-\varepsilon_{ab},
\end{equation}
$G_{ab}$ is the Einstein tensor, $g_{ab}$ is the metric tensor and $\varepsilon_{ab}$ is the electric part of the bulk Weyl curvature. The tensor $S_{ab}$ is given by:
\begin{equation}
S_{ab}=\frac{1}{12}TT_{ab}-\frac{1}{4}T_{ac}T^c_b+\frac{1}{24}g_{ab}(3T_{cd}T^{cd}-T^2),
\end{equation}
Where $T_{ab}$ is the energy-momentum tensor. $\kappa^2$ and $\Lambda$ are the brane gravitational and cosmological constants respectively. The brane tension $\lambda$, bulk cosmological constant $\tilde{\Lambda}$, and bulk gravitational constant $\tilde{\kappa}^2$ are related by: 
\begin{eqnarray}
6\kappa^2 &=&\tilde{\kappa}^2\lambda,\\
2\Lambda &=& \kappa^2\lambda+\tilde{\kappa}^2 \tilde{\Lambda}.
\end{eqnarray}

Negative tension branes have been shown to be unstable \cite{negativebr,mypaper,bigrip}, branes with variable tension have been studied in \cite{brane7}.\par
A different brane-world cosmology proposal has been presented in \cite{brane14} where the bulk black strings (higher-dimensional extensions of black holes) penetrate the brane and lead to a brane with 'Swiss-cheese' structure  \cite{brane15}. The existence of Schwarzschild black holes in such cosmological structure gives rise to some significant cosmological consequences and makes this description more realistic than the FLRW brane \cite{br50}. Such Swiss-cheese brane is different from the old Swiss-cheese proposal suggested by Einstein and Strauss for  embedding compact objects within cosmological space-time. \cite{br51,br52}. The old proposal by Einstein and Strauss is based on removing a spherical comoving region from a FLRW space-time and replacing it by a point mass at its center which creates a hole. Having several non-overlapping spheres where the mass of each sphere contracted to the center (Schwarzschild black holes and surrounding vacuum regions) creates inhomogeneities. The old Einstein-Strauss Swiss-cheese model predicted a modified Luminosity-redshift relation \cite{br53}. \par
A stable flat dark energy-dominated universe in a good agreement with observations has been obtained, in the framework of Swiss-cheese brane-world, in \cite{mypaper} with no crossing to the phantom divide line. The big rip scenario in Swiss-cheese brane-world has been studied in \cite{bigrip} where all the results are independent of the value of the cosmological constant. The evolution of cosmic pressure in \cite{mypaper} shows a sign flipping from positive to negative corresponding to the positive-to-negative sign flipping of the deceleration parameter.

The universe in the Swiss-cheese brane-world scenario is a $4$D FLRW brane embedded in a $5$D Schwarzschild AdS bulk. The FLRW metric is written as
\begin{equation}
ds^{2}=-dt^{2}+a^{2}(t)\left[ \frac{dr^{2}}{1-Kr^2}+r^2d\theta^2+r^2\sin^2\theta d\phi^2 \right] \label{RW}
\end{equation} 
The cosmological equations for a flat universe ($K=0$) are \cite{flat1, flat2, flat3, flat31, flat32}
\begin{eqnarray}
\frac{\dot{a}^2}{a^2} &=& \frac{\Lambda}{3}+\frac{\kappa^2 \rho}{3}\left(1+\frac{\rho}{2\lambda}\right),  \label{cosm1}\\
\frac{\ddot{a}}{a} &=& \frac{\Lambda}{3}-\frac{\kappa^2 }{6}\left[\rho\left(1+\frac{2\rho}{\lambda}\right)+3p \left(1+\frac{\rho}{\lambda}\right) \right] .\label{cosm2}
\end{eqnarray}
GR is recovered for $\rho/\lambda\rightarrow 0$. We will be considering a time-dependent cosmological constant, and here we revise some of the $\Lambda(t)$ models suggested in the literature. A model for $\Lambda(t)$ has been introduced in \cite{21} as
\begin{equation} \label{cosma1}
\Lambda = \frac{\Lambda_{Pl}}{\left(t/t_{Pl}\right)^2} \propto \frac{1}{t^2}
\end{equation}
This begins at the Plank time as $\Lambda_{Pl} = \sim M_{Pl}^2$ and gives rise to the value $\Lambda_{0} \sim 10^{-120} M_{Pl}^2$ at the present epoch. $\Lambda(t)$ decay during inflation had been investigated in \cite{kholopov1,kholopov2}. In terms of the Hubble parameter $H$, the following ansatz has been suggested \cite{20d} where several observations have been used to impose tight restrictions on $\Lambda(t)$ models
\begin{equation} \label{vary2}
\Lambda(H)= \lambda_{0} +\alpha_{0} H + 3 \beta_{0} H^2 
\end{equation}
where $\lambda_{0}$, $\alpha_{0}$ and $\beta_{0}$ are constants. While observations doesn't support the zero value of $\lambda_{0}$ \cite{20b, 20f, 20g, 20h}, $\lambda_{0} \neq 0$ behaves like $\Lambda$CDM model at late-time. Other examples of $\Lambda(H)$ models are \cite{20b}
\begin{eqnarray} 
\Lambda(H)= \beta_{0} H +3H^2 + \delta_{0} H^n,\,\,\,\,\,\, n \in R-\left\{0,1\right\} \\
\Lambda(H, \dot{H}, \ddot{H})=\alpha_{0}+\beta_{0} H+\delta_{0} H^2+ \mu_{0} \dot{H}+\nu_{0} \ddot{H}.
\end{eqnarray} 
A generalized model have been proposed in \cite{Nojiri:2021iko}. The varying $\Lambda$ has also been used in different cosmological contexts such as Bianchi models in $f(R,T)$ gravity and the $4D$ Gauss-Bonnet gravity \cite{n1,n12}.

\section{Scale factor for logamediate inflation }
The logamediate inflation scienario is probed through the scale factor
\begin{equation} \label{annn}
a(t)=a_0 \exp\left(A [\ln t]^{\lambda_1}\right) .
\end{equation}
with $a_0 > 0$, $A > 0$, $\lambda_1 >1$ and $t>1$. This scale factor has been suggested for the first time by Barrow and Nunes in \cite{logam}. They compared the obtained values of the inflationary observables $n_s$ and $r$ to their predicted values
in the intermediate inflation scenario where $a=\exp (Bt^f)$. They also showed that both inflationary scenarios, the logamediate and intermediate, are consistent with observations of the cosmic background radiation. The scale factor represents a form of inflationary expansion in cosmology. Here’s a detailed look at the motivation behind this particular scale factor: (i) Inflationary models in cosmology describe an extremely rapid expansion of the universe in its early stages. The motivation for these models comes from their ability to solve several key problems in the standard Big Bang cosmology, such as the horizon problem, the flatness problem, and the magnetic monopole problem. (ii) In many inflationary models, the scale factor grows exponentially with time. The simplest model is $a(t)= a_{0}e^{H(t)}$, where $H$ is the Hubble parameter. This leads to a constant rate of expansion (de Sitter space). (iii) However, more general forms of inflationary expansion can be considered. The scale factor presented by Eq. (\ref{annn})  falls into this category. This form allows for a richer and more flexible description of the inflationary period, potentially capturing different dynamics and physical processes. (iv) By using a logarithmic term raised to a power, this form allows the growth rate of the scale factor to vary more subtly than a simple exponential. This can accommodate scenarios where the rate of expansion changes over time in a more complex manner than just a constant or linearly changing rate.\\
The Hubble parameter for (\ref{annn}) then is 
\begin{equation} \label{a9n}
H=A \lambda_1 (\ln t)^{\lambda_1-1}\frac{1}{t},
\end{equation}
Which means $A\lambda >0$ for an expanding universe. The Friedmann acceleration equation can be written as
\begin{equation} \label{a9n0}
\frac{\ddot{a}}{a}=\frac{A\lambda_1}{t^2} (\ln t)^{\lambda_1-1} \left[ (\lambda_1-1)(\ln t)^{-1}-1+A\lambda_1 (\ln t)^{\lambda_1-1}\right]
\end{equation}
This scale factor reduces to the power law $a(t)=a_0 t^A$ for $\lambda_1=1$. This scale factor has been used extensively in the literature. It has been shown that although logamediate inflation is not consistent with the {\it Planck 2015} data in the standard framework based on Einstein gravity, it can be compatible with the observational data in the framework of $f(T)$ gravity \cite{ftg}. A model for the Tachyon Logamediate inflation in the framework of RSII braneworld has been constructed in \cite{logam2} where the basic slow-roll parameters were calculated and compared to observations. The predicted values of the spectral index and the tensor-to-scalar fluctuation ratio have been found consistent with those of {\it Planck 2015}. The logamediate inflation driven by a non-canonical scalar field in the context of DGP braneworld gravity has been investigated in \cite{logam3}. The logamediate inflation in the context of Galileon inflation has been studied in \cite{logam4}. A model of warm tachyon inflation in the framework of logamediate inflation has been introduced in \cite{logam5}
 . \par 
Solving equations (\ref{cosm1}) and (\ref{cosm2}) with the varying cosmological constant ansatz (\ref{vary2}) and the scale factor (\ref{annn}) we get
\begin{equation} \label{enn}
\rho=-\frac{1}{2\pi t \ln t } \left[2\pi \lambda t \ln t \pm \left(-\pi \lambda [t^2(\ln t )^2(\lambda_0-4\pi \lambda)+ 3A^2t^2\lambda_1^2(\ln t )^{2\lambda_1}(\beta_0-1)+A\lambda_1 \alpha_0 t (\ln t )^{\lambda_1+1}]\right)^{\frac{1}{2}}\right]
\end{equation}
The $\pm$ means two solutions for $\rho$. We restrict ourselves to the positive tension branes only ( $\lambda>0$ ) where negative tension branes are unstable \cite{negativebr,mypaper,bigrip}. Both solutions in (\ref{enn}) lead to a negative energy density. In the negative sign solution, the energy density $\rho \rightarrow -\infty$ as we get closer to the initial singularity ($t \rightarrow 0 $). In the positive sign solution, the energy density $\rho \rightarrow +\infty$ as $t \rightarrow 0 $. For that reason we stick to the positive sign solution where $\rho$ starts to take negative values only after a finite time from the initial singularity. The domain of the function $\rho(t)$ is restricted to the possible values of the parameters $\alpha_0$, $\beta_0$, $\lambda_0$, $\lambda_1$ and $\lambda$. Since the brane tension $\lambda > 0$, to stay in the real domain of $\rho(t)$ we need $\lambda_0 < 4\pi$ and $\beta_0 < 1$, this guarantees the radicand to be positive if
\begin{equation}
t^2(\ln t )^2+ 3A^2t^2\lambda_1^2(\ln t )^{2\lambda_1} ~>~ A\lambda_1 \alpha_0 t (\ln t )^{\lambda_1+1}
\end{equation}
Then 
\begin{equation}
t^2(\ln t )^2(\lambda_0-4\pi\lambda)+ 3A^2t^2\lambda_1^2(\ln t )^{2\lambda_1}(\beta_0-1) ~<~ A\lambda_1 \alpha_0 t (\ln t )^{\lambda_1+1}, ~~\forall~ \lambda_0 < 4\pi\lambda, ~\beta_0 < 1.
\end{equation}
The pressure can be written as
\begin{eqnarray}
p=-\frac{\lambda}{4t^2\pi(\lambda+\rho)(\ln t )^2}\left[ t^2(\ln t )^2(4\pi \rho+\lambda_0)+A \lambda_1 (\ln t )^{\lambda_1}(\lambda_1-1)+
3A^2\lambda_1^2(\ln t )^{2\lambda_1}(\beta_0-1) \right. \\   \nonumber
\left. +A\lambda_1 (\ln t )^{\lambda_1+1}(\alpha_0t+1)\right] > 0.  
\end{eqnarray}
Inflation is supposed to be driven by a scalar field $\phi$ called inflaton. This is related to the energy density and pressure as
\begin{equation} \label{prho}
\rho_{\phi}=K+V(\phi) ~,~p_{\phi}=K-V(\phi)
\end{equation}
where $K=\frac{1}{2}\dot{\phi}^2$ is the kinetic term and $V=V(\phi)$ is the potential. So, $K=\frac{1}{2}(\rho_{\phi}+p_{\phi})$, $V=\frac{1}{2}(\rho_{\phi}-p_{\phi})$, and $\dot{\phi}^2=(1+\omega_{\phi}) \rho_{\phi}$ with $p_{\phi}=\omega_{\phi} \rho_{\phi}$. The equation of state parameter $\omega_{\phi}$ is defined as
\begin{equation} \label{scc}
\omega_{\phi}= \frac{\frac{1}{2}\dot{\phi}^2-V}{\frac{1}{2}\dot{\phi}^2+V} \simeq -1 ~~~ \text{if} ~~~\frac{1}{2}\dot{\phi}^2 \ll V
	\end{equation}
Both $\rho_{\phi}$ and $p_{\phi}$ are supposed to satisfy the conservation equation
\begin{equation} \label{cons}
\dot{\rho_{\phi}}+3H(\rho_{\phi}+p_{\phi})=0
\end{equation}
Substituting ($\ref{prho}$) into $(\ref{cons})$ leads to the evolution equation for the scalar field
\begin{equation} \label{evol}
\ddot{\phi}+3H\dot{\phi}+V_{,\phi}=0, 
\end{equation}
Where $V_{,\phi}=dV/d \phi$. Here we get the kinetic term and potential as
\begin{equation}
K=-\frac{A\lambda \lambda_1 (\ln t )^{\lambda_1-1}(\ln t-\lambda_1 +1)}{4 t  G(t)} > 0,~~~V(t)=\frac{\lambda F(t)}{4t (\ln t) G(t)} < 0,
\end{equation} 
where 
\begin{eqnarray}
F(t)&=&2t^2(\ln t )^2(\lambda_0-4\pi \lambda)+6A^2\lambda_1^2(\ln t )^{2\lambda_1}(\beta_0-1)+A\lambda_1 (\ln t )^{\lambda_1+1}(2\alpha_0 t+1)\\ \nonumber
&-&4t (\ln t) G(t)+A\lambda_1 (\ln t )^{\lambda_1}(1-\lambda_1)\\  \nonumber
G(t)&=&\left(-\pi \lambda [t^2(\ln t )^2(\lambda_0-4\pi \lambda)+ 3A^2\lambda_1^2(\ln t )^{2\lambda_1}(\beta_0-1)+A\lambda_1 \alpha_0 t (\ln t )^{\lambda_1+1}]\right)^{\frac{1}{2}}.
\end{eqnarray}
The scalar field $\phi(t)$ can be written as
\begin{equation} \label{scalarf}
\phi(t)= \int_0^t \pm \frac{\sqrt{2At\pi^5 \lambda \lambda_1 G(t) (\ln t )^{\lambda_1+1}(\ln t-\lambda_1+1) }}{2\pi^3t (\ln t) G(t)} dt +\phi_0
\end{equation}
Where $\phi_0=\phi(t_0)$. Obtaining $t(\phi)$ to substitute in $V(t)$ and then plot $V(\phi)$ is too difficult for the current model. 
Following the prediction of AdS spaces in string theory and particle physics, negative potential has gained interest in particle physics and cosmology. In super-gravity, string theory, and particle physics, negative potentials are frequently anticipated. They also can be found in cyclic and ekpyrotic cosmological models \cite{cyc,ekp}, the potential of the general vacua of super-gravity is negative. As a result of negative potentials, a high energy scale, such as the electroweak scale or the super-symmetry breaking scale, can be used to explain the cosmic scale \cite{scale}. Scalar field cosmology with negative potentials has been discussed in \cite{linde}. The impact of negative energy densities on the standard FLRW cosmology has been studied in \cite{-ve} where the total energy density $\rho$ can be written as a sum of two power series in terms of the scale factor 
\begin{equation}
\rho=\sum_{n=-\infty}^{\infty}\rho_n^+a^{-n}+\sum_{m=-\infty}^{\infty}\rho_m^-a^{-m},
\end{equation}
where $\rho_n^+$ is the ordinary positive $\rho$, and $\rho_m^-$ is the negative cosmological energy density. It has been also shown that vacuum polarization can represent an example for a gravitational source with negative energy density that could have been a major factor in the early expansion of the universe \cite{-ve2}.
The study in \cite{mexico} obtained an equation of state parameter $\omega_{\phi}<-1$ with no violation of the weak energy condition which requires a negative potential $V(\phi)<0$. It has been also shown that $\rho_{\phi}=\frac{1}{2}\dot{\phi}^2+V(\phi)$ takes negative values with $\omega_{\phi}<-1$, the negative $\rho_{\phi}$ results in a small value of $\Lambda$. Although cosmic expansion exists in such scenario, the universe ends up collapsing because of the negative potential.

\section{ Stability of the model: Energy conditions and causality}
A possible way to examine the physical acceptability of the current model is to test the energy conditions and the sound speed causality condition. For the current Swiss-cheese brane-world model with a quadratic density $\rho^2$-term, the validity of the classical linear energy conditions is not generally expected \cite{parc}. Same situation happens in other modified gravity theories with higher-order curvature corrections, such as Gauss-Bonnet gravity with the Gauss-Bonnet invariant $\mathcal{G} \equiv R^2-4R^{\mu \nu} R_{\mu \nu}+R^{\mu \nu \alpha \beta}R_{\mu \nu \alpha \beta}$, where the classical linear energy conditions are not expected to be valid. It has been shown that
the linear energy conditions (namely, the null $\rho + p\geq 0$; weak $\rho \geq 0$, $\rho + p\geq 0$; strong $\rho + 3p\geq 0$ and dominant $\rho \geq \left|p\right|$) should be replaced by different nonlinear conditions in the existence of semiclassical quantum effects \cite{ec}.\par

Testing those linear conditions shows that both the weak and strong conditions are satisfied ($\rho+p>0$ , and $\rho+3p>0$), only the dominant one is violated ($\rho-p<0$ ). The new nonlinear energy conditions are \cite{ec,FEC1,FEC2}: (i) The flux energy condition $\rho^2 \geq p_i^2$ (ii) The determinant energy condition $ \rho . \Pi p_i \geq 0$ . (iii) The trace-of-square energy condition $\rho^2 + \sum p_i^2 \geq 0$ . We find that 
\begin{eqnarray}
\rho^2-p^2=\frac{(-2\pi \lambda t \ln t +G(t))^2}{4t^2\pi^2(\ln t)^2}- \frac{\lambda^2}{4 t^2(\ln t)^2G^2(t)} \left[t^2(\ln t)^2(\lambda_0-4\pi \lambda)+A\lambda_1 (\ln t )^{\lambda_1}(1-\lambda_1) \right. \\ \nonumber
\left. +3A^2\lambda_1^2 (\ln t )^{2\lambda_1}(\beta_0-1)+A\lambda_1 (\ln t )^{\lambda_1+1}(\alpha_0t+1)+2t\ln t G(t) \right]^2 ~<~0.
\end{eqnarray}
\begin{eqnarray}
\rho~ p^3=\frac{(-2\pi \lambda t \ln t+G(t))\lambda^3}{16 \pi t^4(\ln t)^4G(t)^3} \left[t^2(\ln t)^2(\lambda_0-4\pi \lambda)+A\lambda_1 (\ln t )^{\lambda_1}(1-\lambda_1) \right. \\ \nonumber
\left. +3A^2\lambda_1^2 (\ln t )^{2\lambda_1}(\beta_0-1)+A\lambda_1 (\ln t )^{\lambda_1+1}(\alpha_0t+1)+2t \ln t G(t)\right]^3~>~0
\end{eqnarray}
\begin{eqnarray}
\rho^2+3p^2=\frac{(-2\pi \lambda t \ln t +G(t))^2}{4t^2\pi^2(\ln t)^2} +\frac{3\lambda^2}{4 t^2(\ln t)^2G(t)^2} \left[t^2(\ln t)^2(\lambda_0-4\pi \lambda)+A\lambda_1 (\ln t )^{\lambda_1}(1-\lambda_1) \right. \\ \nonumber
\left. +3A^2\lambda_1^2 (\ln t )^{2\lambda_1}(\beta_0-1)+A\lambda_1 (\ln t )^{\lambda_1+1}(\alpha_0t+1)+2t\ln t G(t) \right]^2 ~>~0.
\end{eqnarray}
Where 
\begin{equation}
G(t)=\sqrt{-\lambda \pi \left( t^2 (\ln t)^2 (\lambda_0-4\pi\lambda ) +At\alpha_0\lambda_1(\ln t)^{\lambda_1+1} + A^2 \lambda_1^2(\ln t)^{2\lambda_1} (3\beta_0-2) \right)}
\end{equation}

So, we now have the situation where $\rho < 0$ with $p < 0$ lead to $+K$, $-V$, and the validity of the two conditions $\rho^2 + \sum p_i^2 \geq 0$  and  $ \rho . \Pi p_i \geq 0$. The sound speed causality condition $0 \leq \frac{dp}{d\rho} \leq 1$ is not satisfied. it's superluminal ($\frac{dp}{d\rho} > 1$) at early-times and negative ($\frac{dp}{d\rho} < 0$) at late-times. The energy conditions have been plotted in Fig. (1) for $A=5$, $\lambda_0=\alpha_0=1$, $\beta_0=0.5$, $\lambda_1=1.2$ and $\lambda=0.1$.
\begin{figure}[H] \label{99}
  \centering     
					  \subfigure[$$]{\label{F1009}\includegraphics[width=0.35\textwidth]{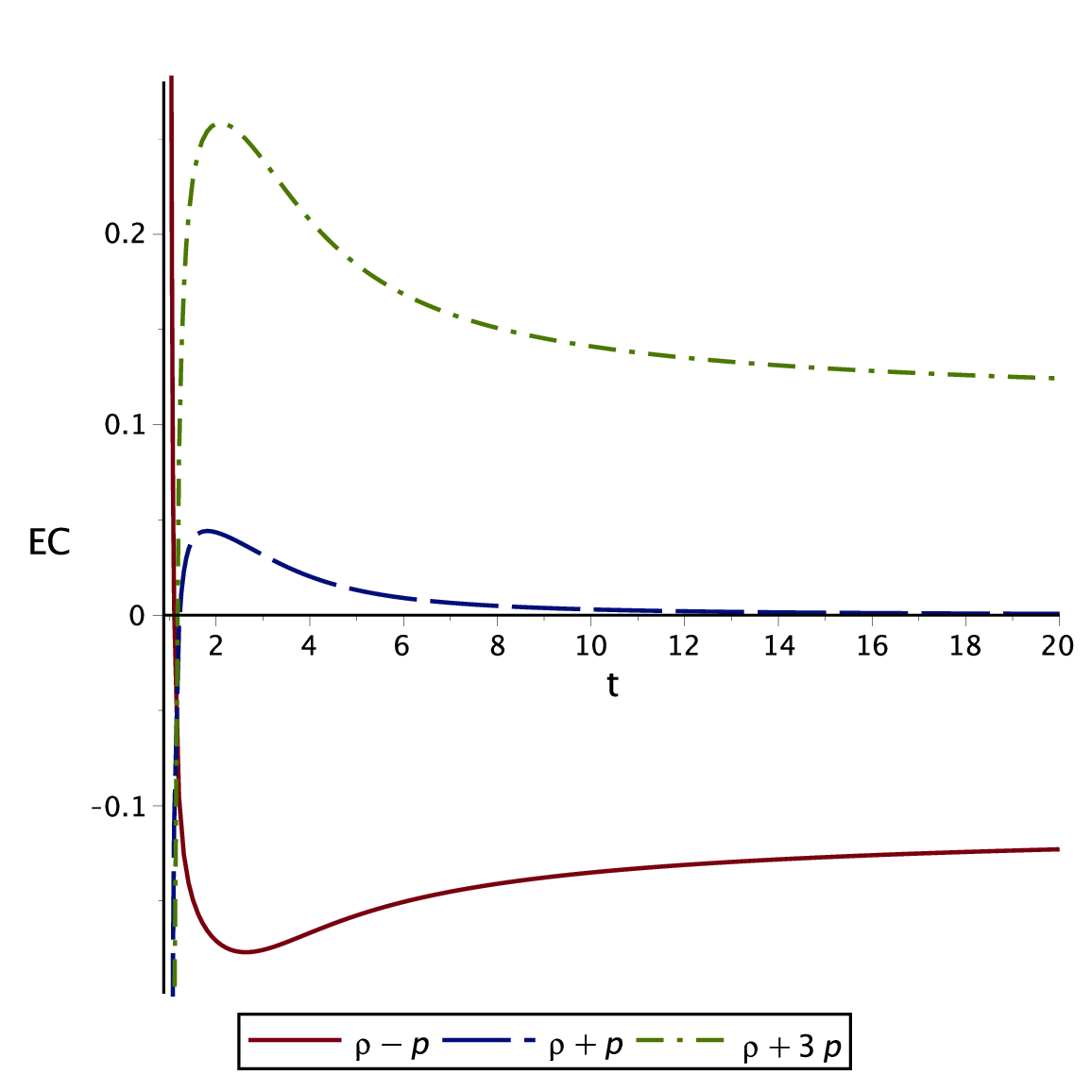}} \hspace{8mm} 
	  \subfigure[$$]{\label{0900}\includegraphics[width=0.35\textwidth]{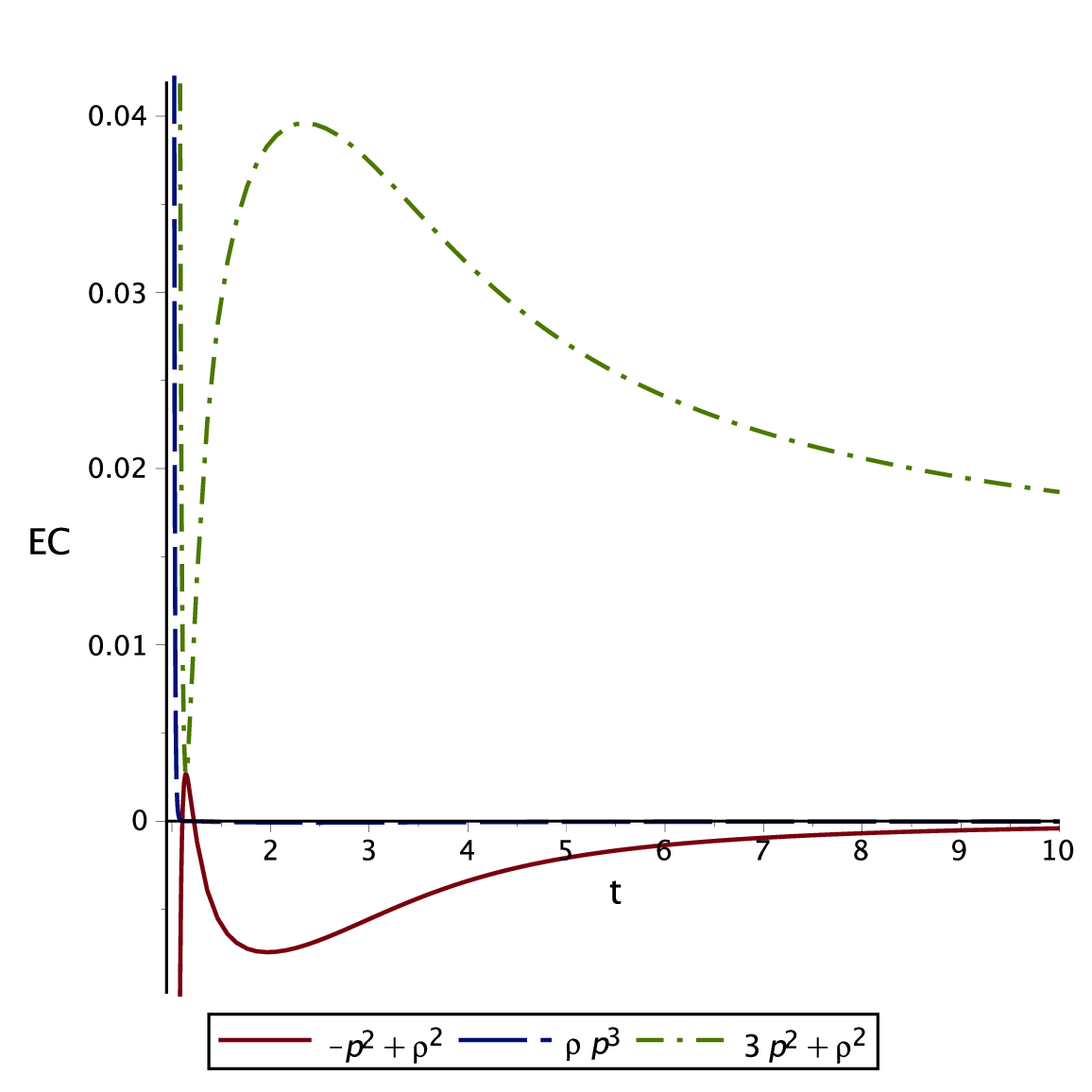}} \hspace{8mm} 
  \caption{ The validity of energy conditions for the current model (a) Linear ECs. (b) Nonlinear ECs}
  \label{fig:01}
\end{figure}
\section{Slow-roll parameters and eternal inflation problem}
The inflation era is not intended to last forever and it has to come to an end. In the same time, it should last enough to solve the initial conditions' problem which means a sufficient number of efoldings $N\approx 55-65$ is required. While the slow-roll parameters for the current inflationary model can be easily investigated, there is nothing to stop the inflation (an eternal inflation). This has been noted before for logamediate inflation in \cite{ira}. The inflation parameters are all expressed in terms of the Hubble parameter based on the ad hoc scale factor we have assumed. We have not obtained the scale factor by solving the Swiss-cheese brane-world cosmological equations (\ref{cosm1})  and (\ref{cosm2}). Therefore, the inflationary parameters obtained here are all model-independent and we don't expect them to reflect the inhomogeneities of the Swiss-cheese brane. The slow-roll dimensionless parameters are given as
\begin{equation}\label{epss}
\epsilon=-\frac{\dot{H}}{H^2}=\frac{\ln t -\lambda_1+1}{A\lambda_1 (\ln t)^{\lambda_1}}
\end{equation}
\begin{equation}
\eta=-\frac{1}{2}\frac{\ddot{H}}{\dot{H}H}=\frac{2(\ln t)^2+3(\ln t) (1-\lambda_1)+\lambda_1^2-3\lambda_1+2}{2A\lambda_1 (\ln t)^{\lambda_1}(\ln t -\lambda_1+1)}
\end{equation}
The spectral index for curvature perturbations is 
\begin{equation}
n_s=1-6\epsilon+2\eta=-\frac{2(\ln t)^2+6(\ln t) (1-\lambda_1)+4\lambda_1^2-6\lambda_1+2}{2A\lambda_1 (\ln t)^{\lambda_1}(\ln t -\lambda_1+1)}+1
\end{equation}
The tensor to scalar ratio is given by:
\begin{equation}
r=16 \epsilon=16\frac{\ln t -\lambda_1+1}{A\lambda_1 (\ln t)^{\lambda_1}}
\end{equation}
Equation (\ref{epss}) shows that $\epsilon$ starts to increase at $t=1$ and keeps increasing until it reaches the maximum value $\epsilon_{max}$ at a specific time $t_{max}$. Then, it starts to decrease and approaches zero as $t\rightarrow \infty$.  We also know that inflation takes place for $\ddot{a}>1$ which is proportional to $\epsilon <1$, so that $\epsilon =1$ at the start or at the end of the inflationary era. Therefore, we are interested only in the situations with the condition $\epsilon_{max}\geq 1$. This condition leads to the constraint \cite{logam3} $A\leq \lambda_1^{-\lambda_1 -1}$. It has also been shown by Barrow \& Nunes \cite{logam} that the approximation $\ln t \gg \lambda-1$ is satisfied at late-times. This leads to $\epsilon =(\ln t)^{1-\lambda}/(A\lambda)$ and we get $t_b=\exp [(A\lambda_1)^{\frac{1}{1-\lambda_1}}]$.\newline
The number of efoldings is
\begin{equation}
N=\int_{t_1}^{t_2} H dt=A[(\ln t_2)^{\lambda_1}-(\ln t_1)^{\lambda_1}]\approx 55 - 65
\end{equation}
Following the analysis given in \cite{logam3}, and cosidering $t_1=t_b$ we find
\begin{equation}\label{lnt}
\ln t = \left[\frac{N}{A}+(A\lambda_1)^{\frac{\lambda_1}{1-\lambda_1}}\right]^{1/\lambda_1}
\end{equation}
We now can express the inflation parameters in terms of the number of efoldings $N$. We have plotted $\epsilon(N)$, $\eta(N)$ and $n_s(N)$ in the following Figure 
\begin{figure}[H] \label{96}
  \centering     
					  \subfigure[$\epsilon(N)$]{\label{F5t}\includegraphics[width=0.3\textwidth]{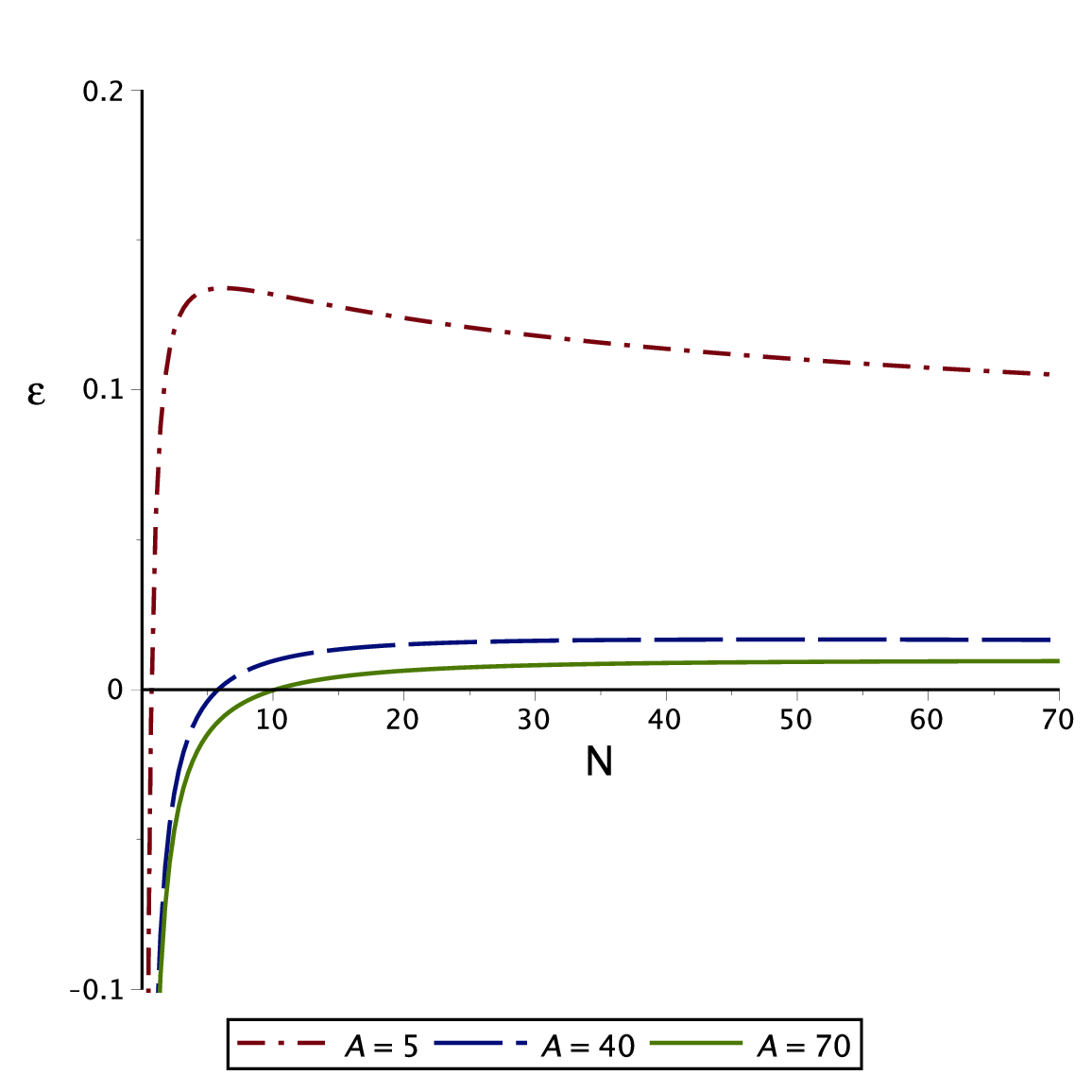}} \hspace{4mm} 
	  \subfigure[$\eta(N)$]{\label{yt0}\includegraphics[width=0.3\textwidth]{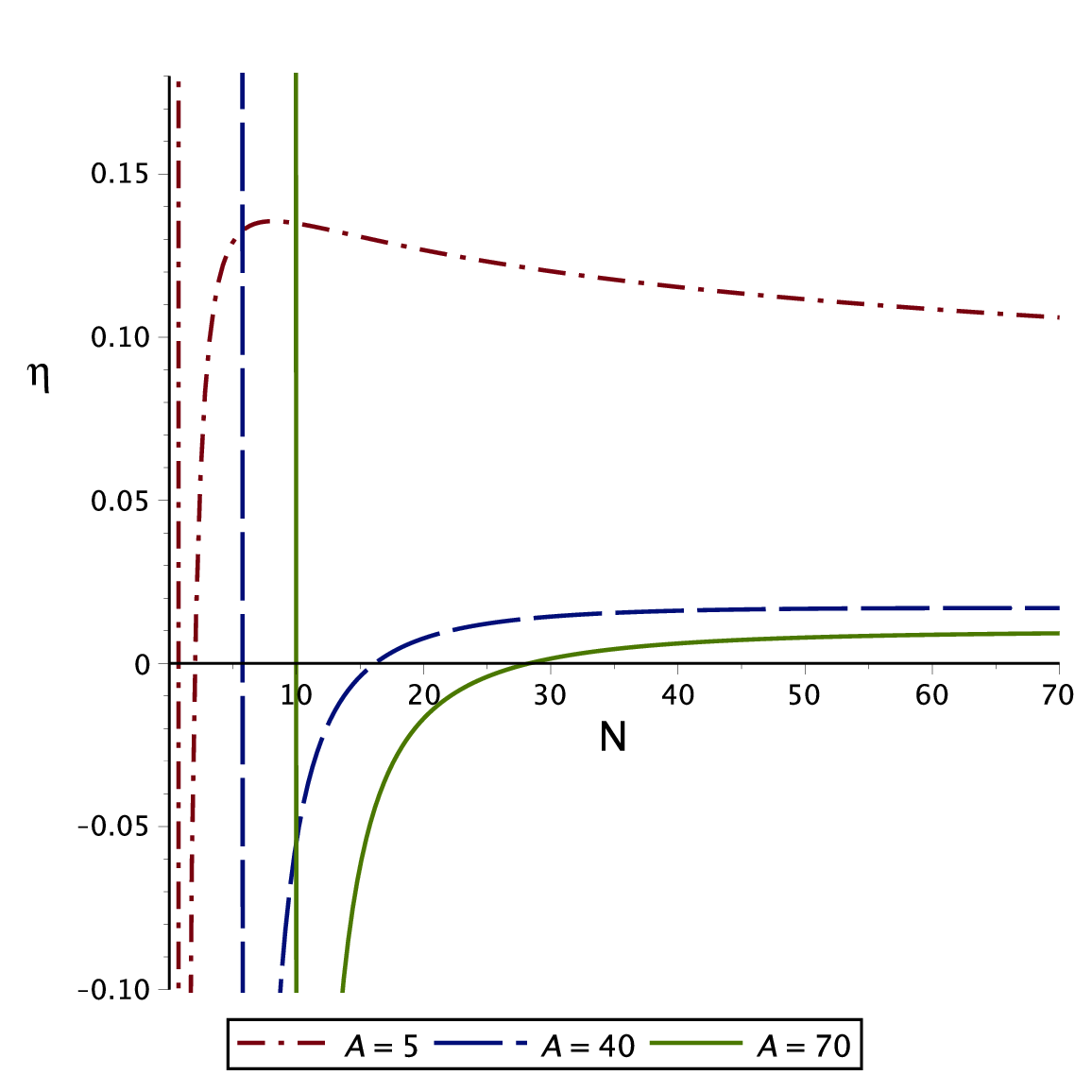}} \hspace{4mm} 
		 \subfigure[$n_s(N)$]{\label{rt0}\includegraphics[width=0.3\textwidth]{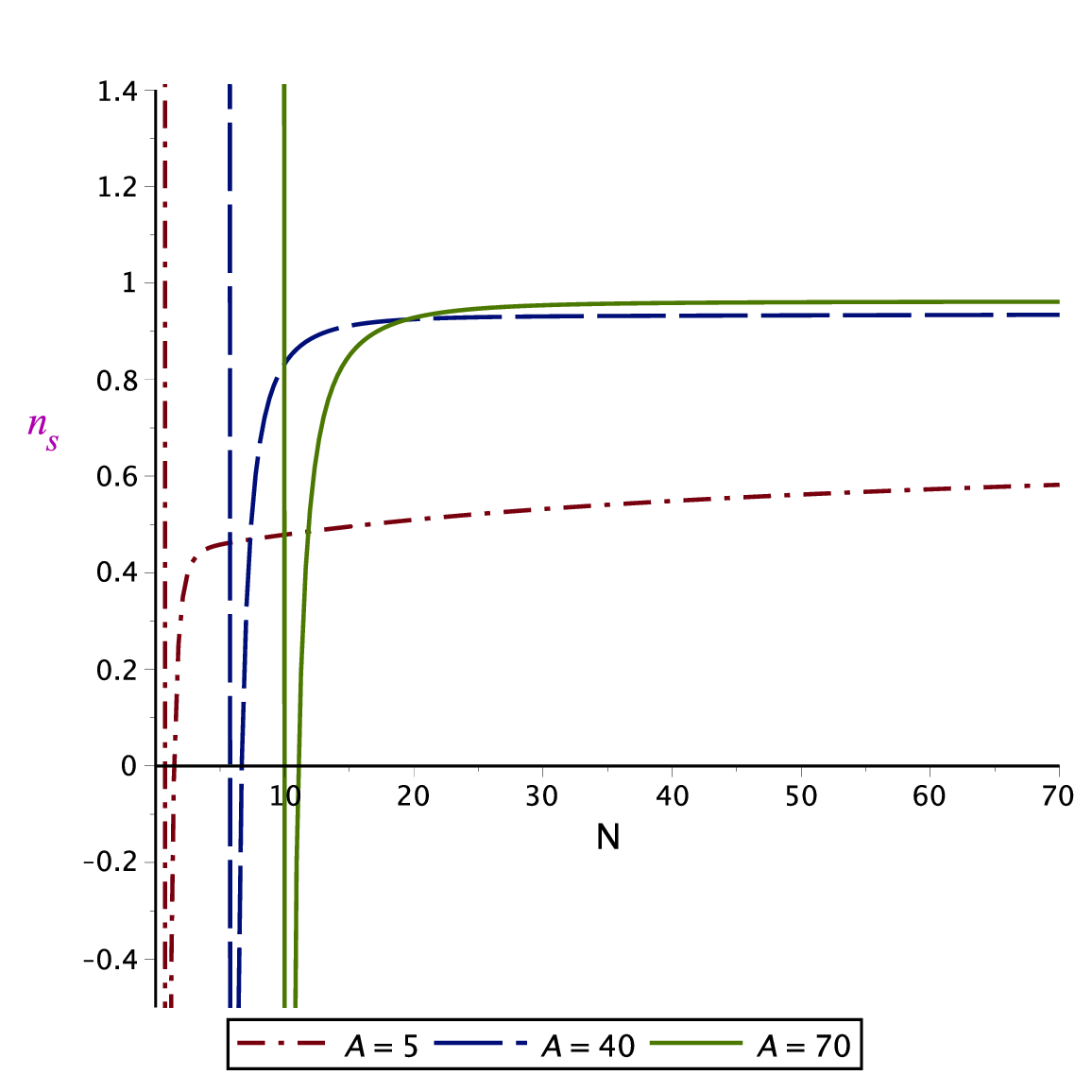}} \hspace{4mm} 
  \caption{ The evolution of the inflation parameters as functions of the number of efoldings $N$ plotted for several values of $A$. $n_s$ agrees with Planck 2015 results for $50 \lesssim A \lesssim 80 $.}
  \label{fig:32}
\end{figure}
The analysis of Planck data sets has been performed in \cite{plank1,plank2}. The analysis shows that the single scalar field inflationary models in slow-roll limit have limited spectral index, very low tensor-scalar ratio $n_s=0.968\pm 0.006$ and $r < 0.11$. The evolution of $n_s$against $N$ shows a good agreement with Planck 2015 results (Fig.2(c)) for $50 \lesssim A \lesssim 80 $.
For the first slow-roll parameter $\epsilon$, we can easily calculate the limit as $t$ tends to infinity
\begin{equation}
\lim_{t \rightarrow \infty}  \frac{1}{A\lambda_1}(\ln t)^{-\lambda_1}(\ln t -\lambda_1+1)=0 
\end{equation}
So $\epsilon$ begins to increase at $t=1$ and continues to increase until a maximum value, then it starts to decrease until it reaches $0$ at $+\infty$. Since inflation ends when $\epsilon \approx 1$, in the current model $\epsilon$ is a decreasing function at late times ($t \gg 1$) and it can't approaches $1$. So inflation will never end and the inflaton scalar field holds for late-times too not only for early-times. Some attempts have been suggested in the literature to solve the problem of eternal acceleration. For example, It has been suggested in \cite{negativel} that the existence of negative cosmological constant $\Lambda < 0$ solves the eternal inflation problem. However, this suggested non-conventional solution with a negative cosmological constant can't be checked for the current model as the first slow-roll parameter $\epsilon$ can't be expressed as a function of the cosmological constant $\Lambda$. 

\section{Cosmography}

The cosmographic analysis approach allows cosmological parameters to be expressed only through kinematics \cite{cosmography1,cosmography2, cosmoref1,cosmoref2,cosmoref3,cosmoref4,cosmoref5,cosmoref6,cosmoref7,cosmoref8}. This analysis is therefore independent of the specific model under consideration and there is no need for EoS to study the dynamics \cite{cosmography3}. Although the cosmographic approach is considered as a model-independent technique able to describe the late-time cosmic expansion, relating  cosmography to the early universe has been considered by some authors \cite{cosmog3,cosmog4}. The cosmographic nature of the early universe from extra dimensional viewpoint has been discussed in \cite{cosmog3}. In \cite{cosmog4}, the inflationary era has been investigated with slow-roll approximation in the context of cosmography parameters and the slow-roll parameters were determined in terms of cosmographic parameters. Then, the spectral index, tensor to scalar ratio and number of e-foldings have also been expressed as functions of cosmographic parameters. The early and late epochs have been connected by the suggested $f(z)$CDM cosmography in \cite{cosmog5}. The scale factor $a(t)$ can be expanded around the present time $t_0$ as
\begin{equation} \label{taylor}
a(t)=a_0 \left[ 1+ \sum_{n=1}^{\infty} \frac{1}{n!} \frac{d^na}{dt^n} (t-t_0)^n \right]
\end{equation}
The following coefficients are denoted as the Hubble $H$, deceleration $q$, jerk $j$, snap $s$, lerk $l$ and max-out $m$ parameters respectively
\begin{eqnarray} 
H=\frac{1}{a}\frac{da}{dt}~~,~~q=-\frac{1}{aH^2}\frac{d^2a}{dt^2}~~,~~j=\frac{1}{aH^3}\frac{d^3a}{dt^3}\\   \nonumber
s=\frac{1}{aH^4}\frac{d^4a}{dt^4}~~,~~l=\frac{1}{aH^5}\frac{d^5a}{dt^5}~~,~~m=\frac{1}{aH^6}\frac{d^6a}{dt^6}.
\end{eqnarray} 
The derivatives of $H$ can also be related to other cosmographic parameters \cite{cosmo11}
\begin{eqnarray*} 
\dot{H}&=&-H^2 (1+q)~~,~~\ddot{H}=H^3 (j+3q+2)~~,~~\frac{d^3H}{dt^3}=H^4[s-4j-3q(q+4)-6],\\
\frac{d^4H}{dt^4}&=&H^5[l-5s+10(q+2)(j+3q)+24].
\end{eqnarray*} 
To set restrictions on the cosmographic expansion, a high-redshift study up to the fifth order has been carried out in \cite{cosmo11} where the present accelerating cosmic expansion \cite{new1,new2} has been confirmed. For the present model we get
\begin{eqnarray}
j(t)&=&\frac{1}{A^2\lambda_1^2}\left[(-3\lambda_1+3)(\ln t)^{-2\lambda_1+1}+2(\ln t)^{-2\lambda_1+2}-3A\lambda_1 (\ln t)^{-\lambda_1+1} \right. \\ \nonumber
&+& \left.(\lambda_1^2+3\lambda_1+2)(\ln t)^{-2\lambda_1}+(3A\lambda_1^2-3A\lambda_1)(\ln t)^{-\lambda_1}+A^2\lambda_1^2\right].\\ 
q(t)&=&{\frac { \left( -\lambda_1+1 \right) (\ln t)^{-\lambda_1}-A\lambda_1+ (\ln t)^
{-\lambda_1+1}}{A\lambda_1}}.
\end{eqnarray}
\begin{eqnarray}
s&=&\frac{1}{A^3\lambda_1^3}\left[A\lambda_1 (\ln t)^{-2\lambda_1}(7\lambda_1^2-18\lambda_1+11)+18A\lambda_1^2 (\ln t)^{-2\lambda_1+1}(1-\lambda_1)\right.\\    \nonumber
&+& \left. 6A^2\lambda_1^2 (\ln t)^{-\lambda_1}(\lambda_1-1) + 11A\lambda_1 (\ln t)^{-2\lambda_1+2}-6A^2\lambda_1^2 (\ln t)^{1-\lambda_1}  \right.\\     \nonumber
&-& 6(\ln t)^{-3\lambda_1+1} (\lambda_1^2-3\lambda_1+2) +\left.(\ln t)^{-3\lambda_1}(\lambda_1^3-6\lambda_1^2+11\lambda_1-6)\right.\\     \nonumber
&+&\left. 11\lambda_1 (\ln t)^{-3\lambda_1+2}(\lambda_1-1)-6(\ln t)^{-3\lambda_1+3}\right].
\end{eqnarray}
\begin{eqnarray}
l&=&\frac{1}{A^4\lambda_1^4}\left[A^4\lambda_1^4-10(\ln t)^{-4\lambda_1+1}(\lambda_1^3-6\lambda_1^2+11\lambda_1-6)+35 (\ln t)^{-4\lambda_1+2}(\lambda_1^2-3\lambda_1+2)\right.\\     \nonumber
&-&\left.50(\ln t)^{-4\lambda_1+3}(\lambda_1-1)+24(\ln t)^{-4\lambda_1+4}-10A^3\lambda_1^3(\ln t)^{1-\lambda_1}+60A^2\lambda_1^3(\ln t)^{-2\lambda_1+1}(1-\lambda_1)\right.\\     \nonumber
&+&\left. 35A\lambda_1 (\ln t)^{-2\lambda_1+2}(A\lambda_1+3\lambda_1-3)-10(\ln t)^{-3\lambda_1+1}(7A\lambda_1^3-18A\lambda_1^2+11)\right.\\     \nonumber
&+&\left.105(\ln t)^{-3\lambda_1+2}(A\lambda_1^2-1) -50A\lambda_1(\ln t)^{-3\lambda_1+3}+5A(\ln t)^{-3\lambda_1}(3\lambda_1^4-7\lambda_1^3+21\lambda_1^2-10\lambda_1) \right.\\     \nonumber
&+&\left. 5A^2(\ln t)^{-2\lambda_1}(5\lambda_1^4-12\lambda_1^3+7\lambda_1^2)+10A^3\lambda_1^3(\ln t)^{-\lambda_1}(\lambda_1-1)  \right.\\     \nonumber
&+&\left.(\ln t)^{-4\lambda_1}(\lambda_1^4-10\lambda_1^3+35\lambda_1^2-50\lambda_1+24)  \right].
\end{eqnarray}
\begin{eqnarray}
m &=& \frac{1}{A^5\lambda_1^5} \left[ A\lambda_1(\ln t)^{-4\lambda_1}(31\lambda_1^4-225\lambda_1^3+595\lambda_1^2-675\lambda_1+274)  \right.\\     \nonumber
  &+& \left. 5A^2\lambda_1^2(\ln t)^{-3\lambda_1}(18\lambda_1^3-75\lambda_1^2+105\lambda_1-45)+5A^3\lambda_1^3 (\ln t)^{-2\lambda_1}(13\lambda_1^2-30\lambda_1+17) \right.\\     \nonumber
 &+& \left.15A^4\lambda_1^4(\ln t)^{-\lambda_1}(\lambda_1-1)+\lambda_1(\ln t)^{-5\lambda_1}(\lambda_1^4+15\lambda_1^3+85\lambda_1^2-225\lambda_1+274)\right.\\     \nonumber	
&+& \left. 15(\ln t)^{-5\lambda_1+1}(-\lambda_1^4+10\lambda_1^3-35\lambda_1^2+50\lambda_1+24)-120(\ln t)^{-5\lambda_1+5}\right.\\     \nonumber
&+& \left. 274\ln t)^{-5\lambda_1+4}-450(\ln t)^{-5\lambda_1+3}+85(\ln t)^{-5\lambda_1+2}(\lambda_1^3-6\lambda_1^2+11\lambda_1-6) \right.\\     \nonumber
&-& \left.15 A^4\lambda_1^4(\ln t)^{1-\lambda_1}+ 150A^3\lambda_1^3 (\ln t)^{-2\lambda_1+1}(1-\lambda_1)+85A^3\lambda_1^3 (\ln t)^{-2\lambda_1+2} +\right.\\     \nonumber
 &+& \left. 510A^2\lambda_1^2 (\ln t)^{-3\lambda_1+2}-225A^2\lambda_1^2 (\ln t)^{-3\lambda_1+3} -75A^2\lambda_1^2 (\ln t)^{-3\lambda_1+1} (5\lambda_1^2-12\lambda_1+7) \right.\\     \nonumber
&-& \left.  675A\lambda_1 (\ln t)^{-4\lambda_1+3}(\lambda_1-1)+274A\lambda_1 (\ln t)^{-4\lambda_1+4}+85A\lambda_1 (\ln t)^{-4\lambda_1+2}(7\lambda_1^2-18\lambda_1+11)   \right.\\     \nonumber
&-& \left.  75A\lambda_1 (\ln t)^{-4\lambda_1+1}(3\lambda_1^3-14\lambda_1^2+21\lambda_1-10)+85(\ln t)^{-5\lambda_1+8}(\lambda_1^3+11\lambda_1-6) \right.\\     \nonumber
&-& \left.  225\lambda_1 (\ln t)^{-5\lambda_1+3}(\lambda_1-3)+274\lambda_1 (\ln t)^{-5\lambda_1+4}-120(\ln t)^{-5\lambda_1}\right.\\     \nonumber
&-& \left.  15 \lambda_1 (\ln t)^{-5\lambda_1+1}(\lambda_1^3-10\lambda_1^2+35\lambda_1-50)+A^5\lambda_1^5\right].
\end{eqnarray}

\begin{figure}[H] \label{tap1}
  \centering     
					  \subfigure[$$]{\label{F19}\includegraphics[width=0.35\textwidth]{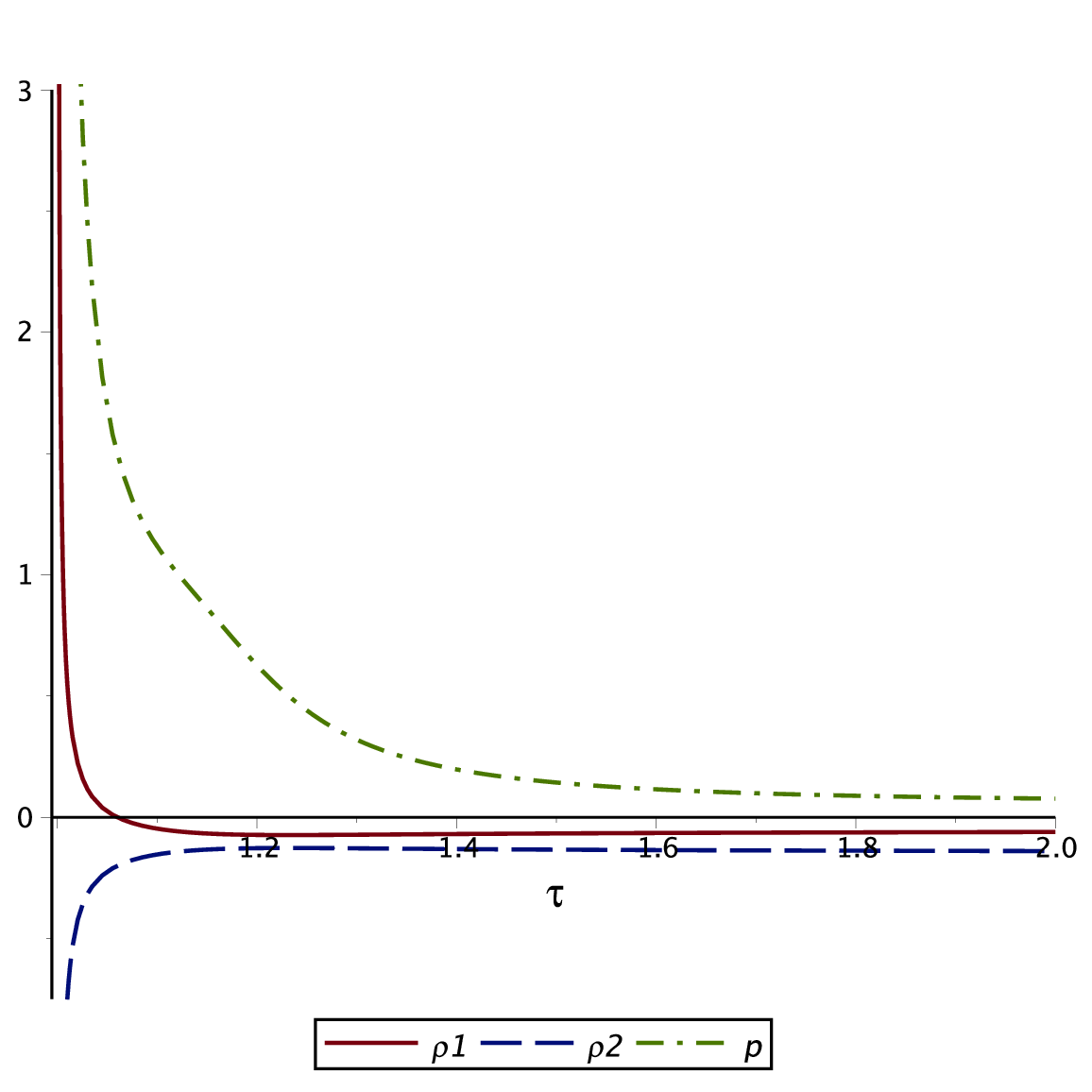}} \hspace{8mm} 
	  \subfigure[$$]{\label{F1900}\includegraphics[width=0.35\textwidth]{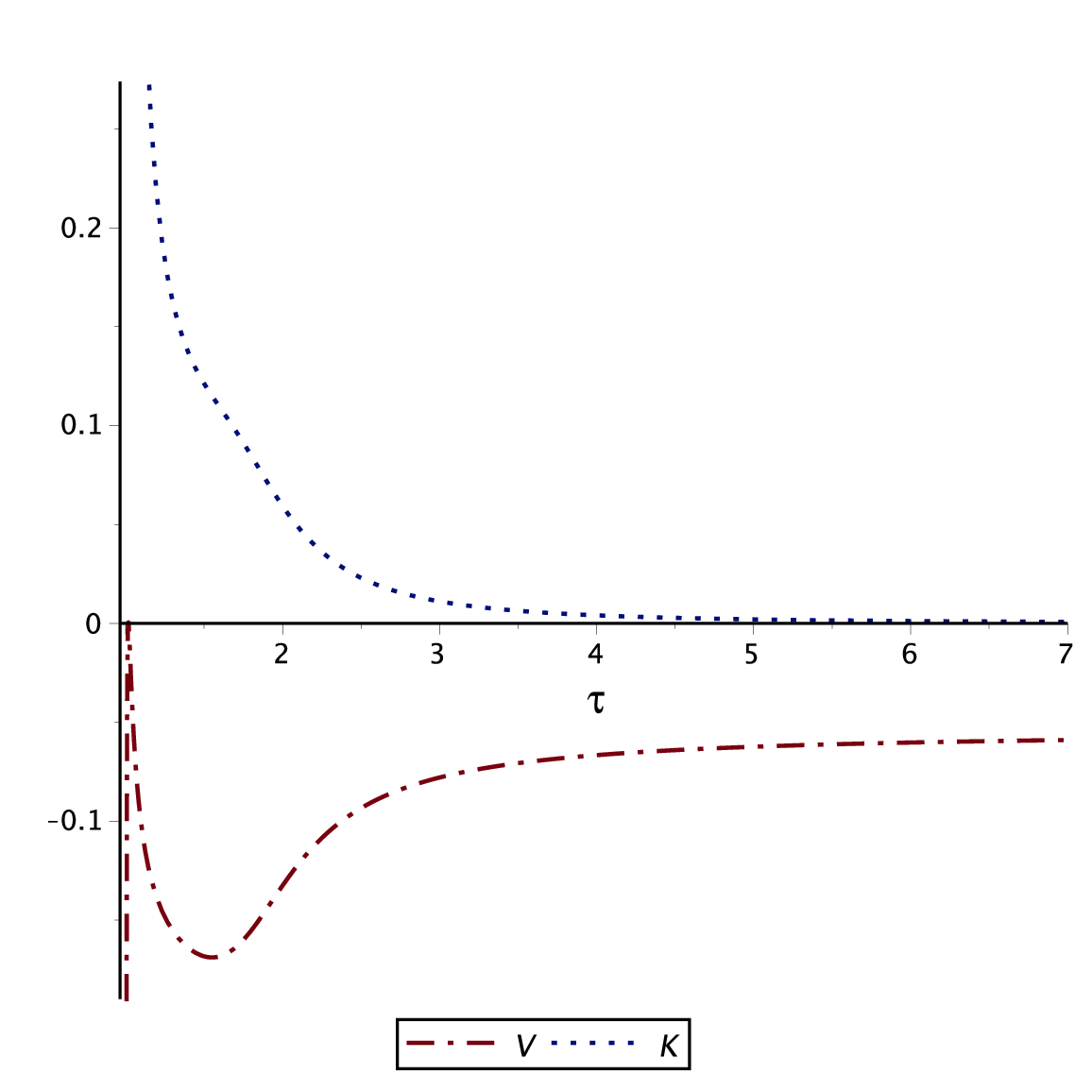}} \hspace{8mm} \\
  \subfigure[$$]{\label{F1}\includegraphics[width=0.35\textwidth]{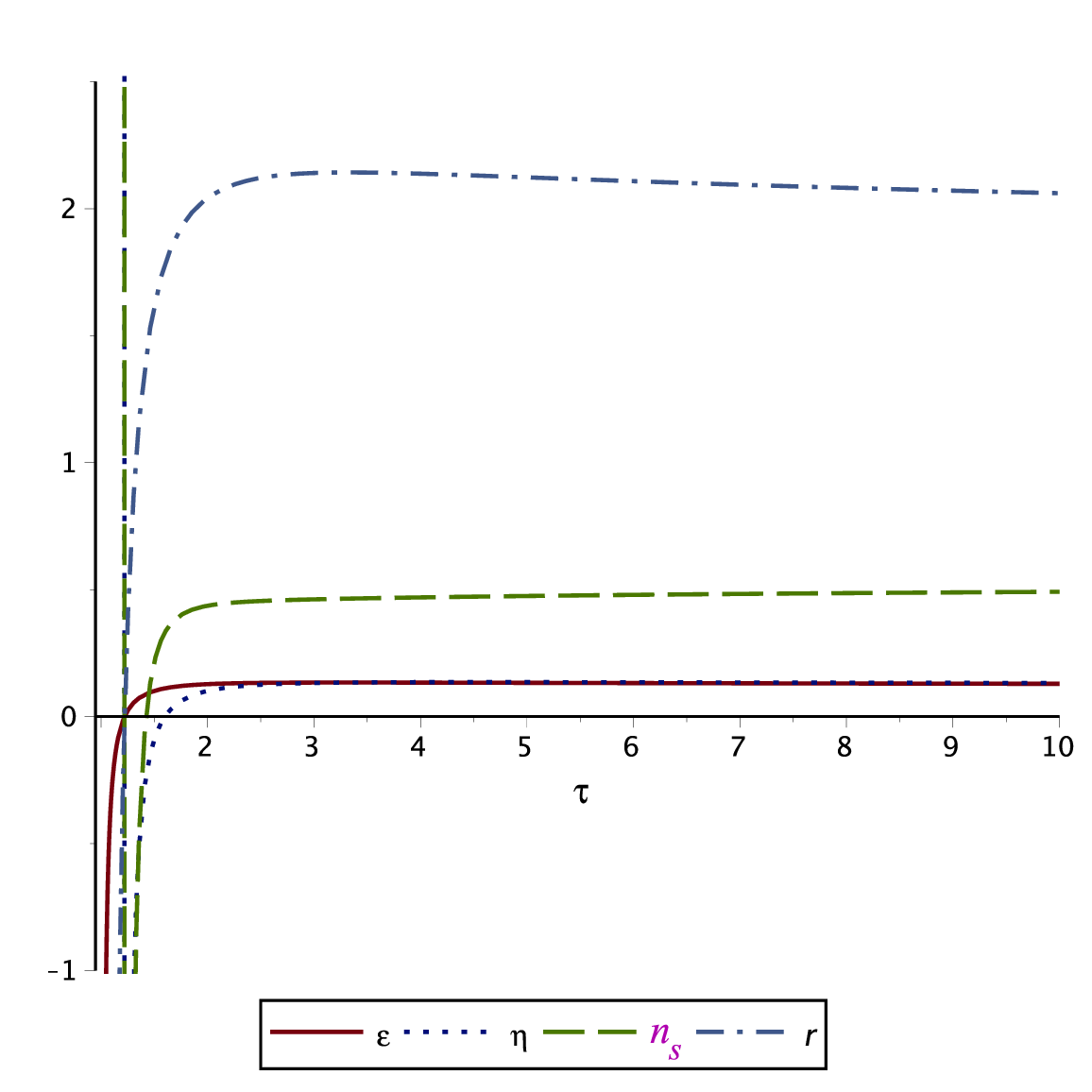}} \hspace{8mm} 
			  \subfigure[$$]{\label{F4}\includegraphics[width=0.35 \textwidth]{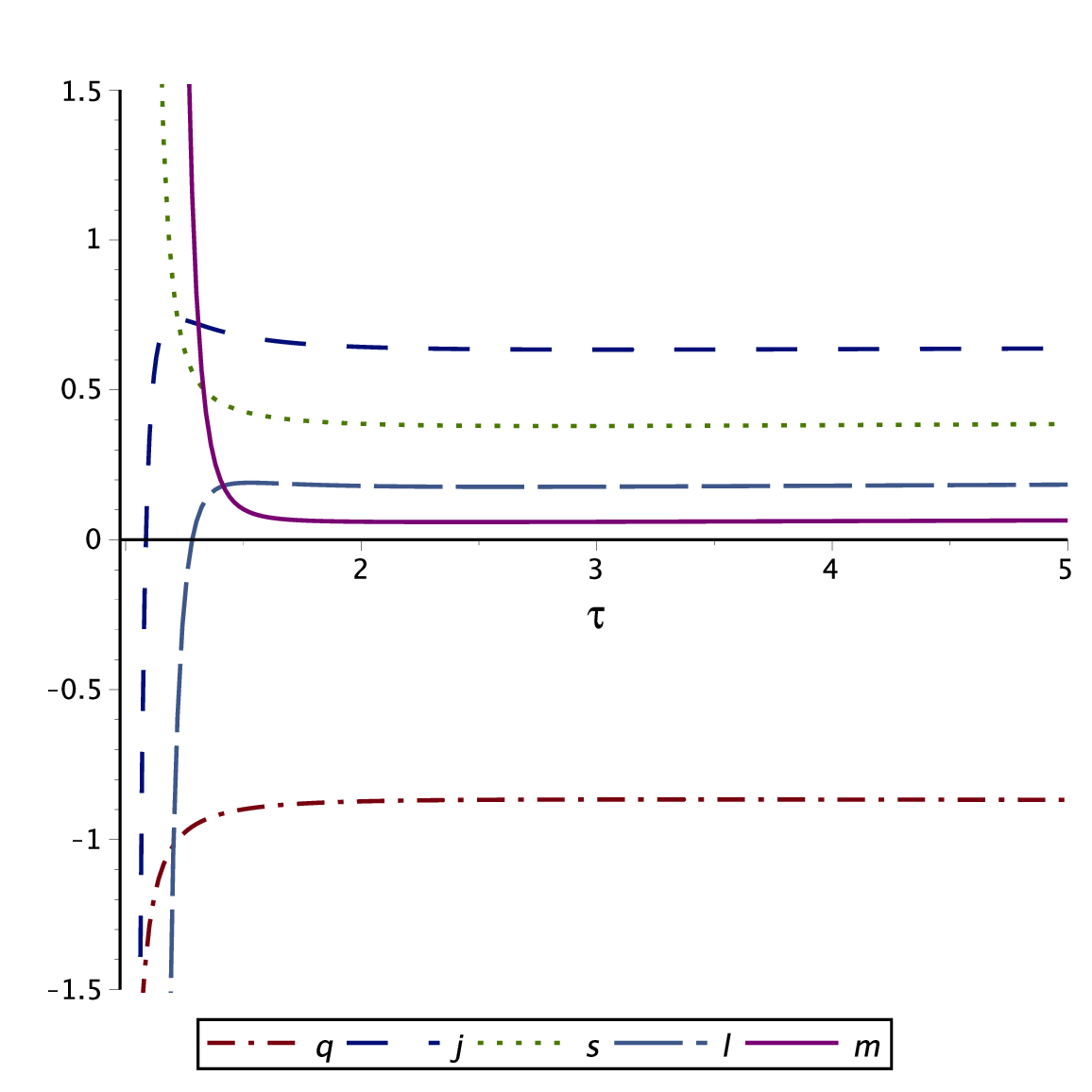}}\\
	\subfigure[$$]{\label{F48}\includegraphics[width=0.35 \textwidth]{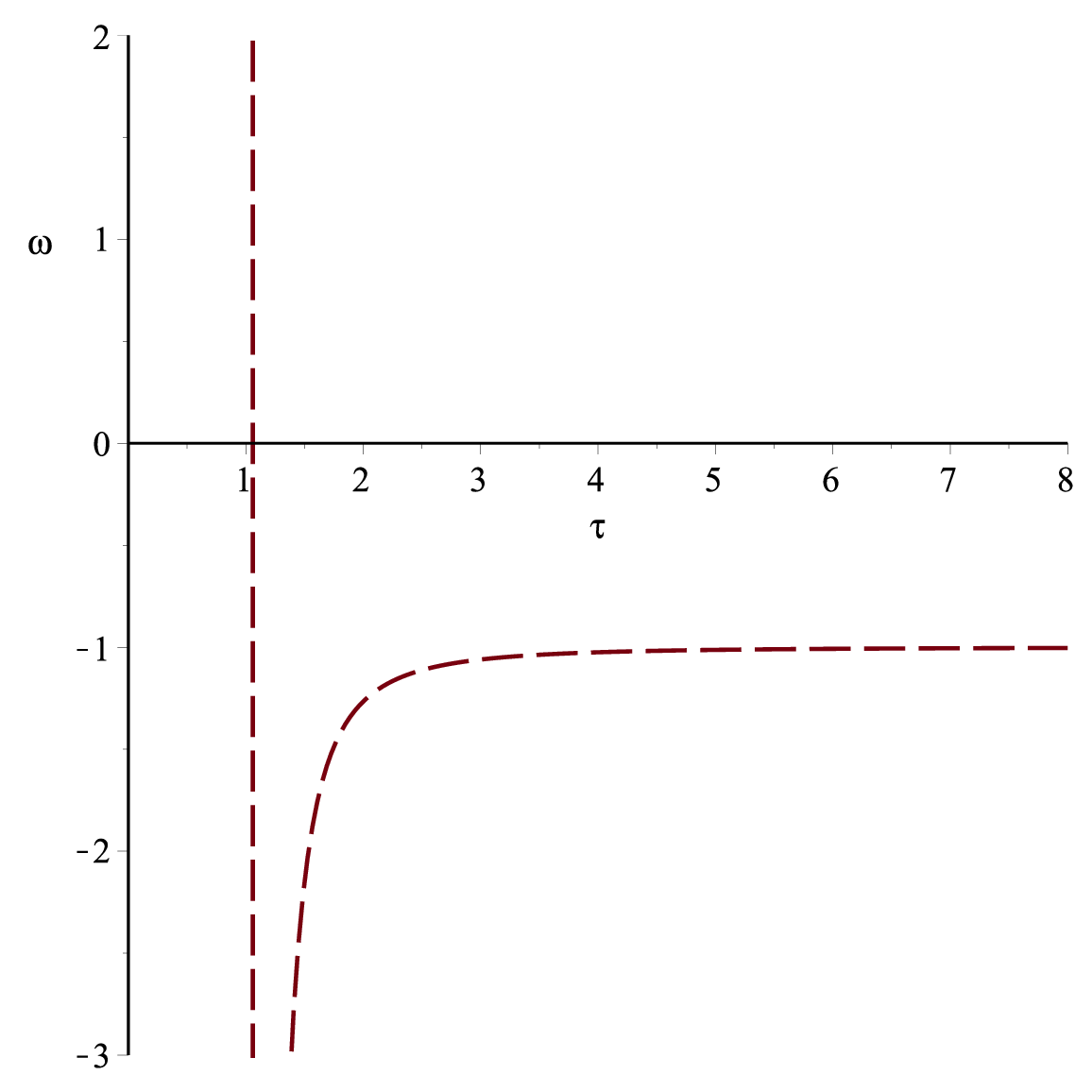}}		
	\subfigure[$$]{\label{F408}\includegraphics[width=0.35 \textwidth]{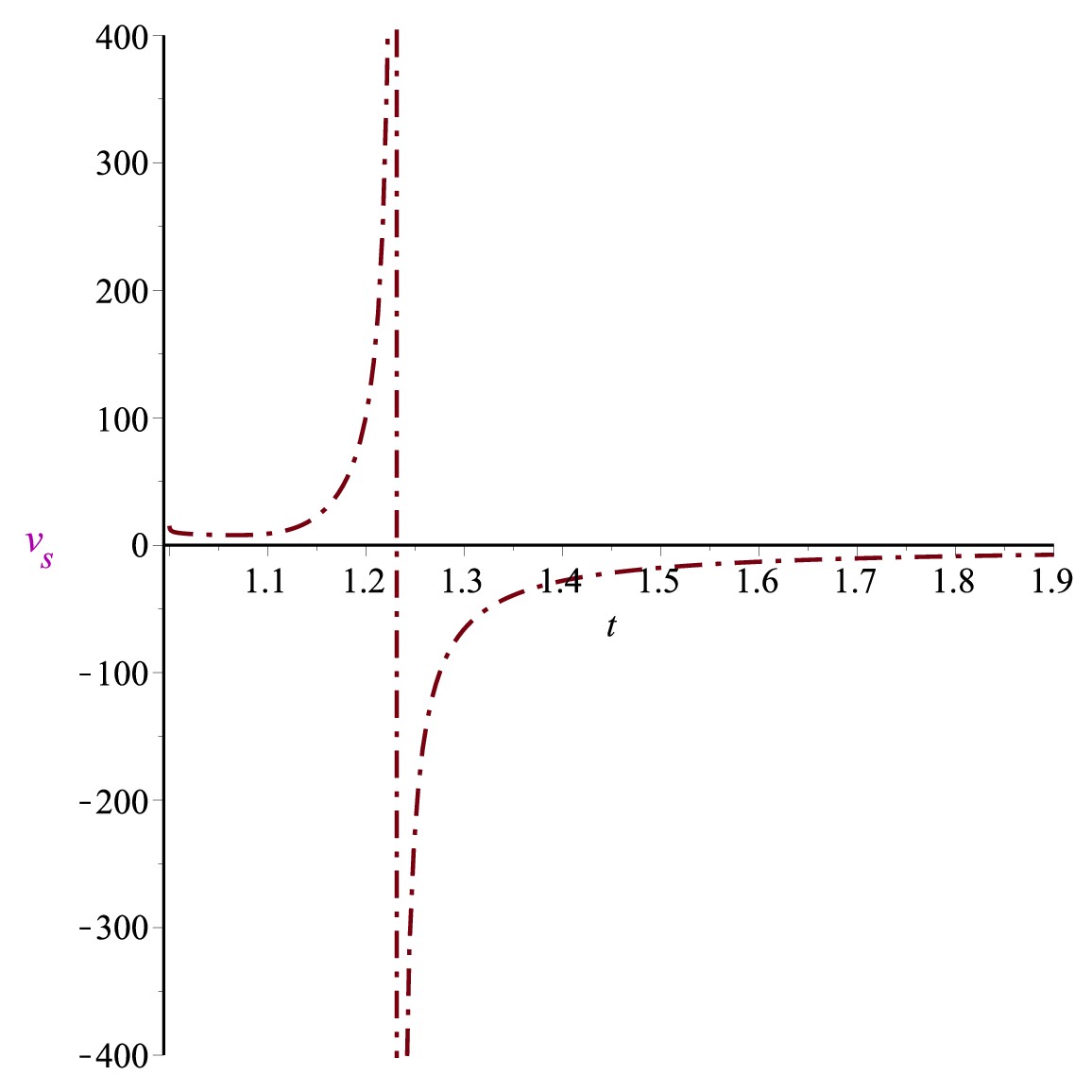}}		
  \caption{ (a) The two solutions for $\rho$, we choose the first one which is positive at early-times and tends to $+\infty$ near the initial singularity. There is a time interval where both $\rho$ and $p$ are positive. (b) A positive kinetic term and a negative potential with AdS minimum. (c) \& (d) slow-roll and cosmographic parameters. (e) The EoS parameter $\omega \rightarrow -1$ for the present era and late-times. (f) The sound speed causality condition $0 \leq \frac{dp}{d\rho} \leq 1$ is not satisfied, it's superluminal ($\frac{dp}{d\rho} > 1$) at early-times and negative ($\frac{dp}{d\rho} < 0$) at late-times.}
  \label{fig:1}
\end{figure}
\section{Conclusion}
The current work explores the Logamediate inflation scenario in the framework of the Swiss-cheese brane-world with varying cosmological constant given by $\Lambda(H)= \lambda_{0} +\alpha_{0} H + 3 \beta_{0} H^2 $. We have investigated the behavior of energy density, pressure, EoS parameter, kinetic term and potential. Although positive energy dominates the present expanding universe, the possibility for the existence of a stable negative energy form whose current density is negligible but could dominate the cosmic future has been discussed in \cite{-ve}. As this kind of role could only be filled by a negative phantom energy (where $\omega<-1$), the current model doesn't support such future scenario where we obtain $\omega \rightarrow -1$ for the present era and late-times. We have obtained a negative potential with AdS minimum and a positive Kinetic term. The pressure is positive $p > 0$ and the energy density $\rho \rightarrow +\infty$ as $\tau \rightarrow  0$. While $\rho$ starts to be negative after a finite time, there is a time interval where both $\rho$ and $p$ are positive.\par

The model suffers from the eternal inflation problem which appears from the evolution of the first slow-roll parameter $\epsilon$. Expressing the logamediate inflation parameters in terms of the number of efoldings $N$ shows a good agreement with Planck 15 observations for the spectral index for curvature perturbations $n_s$ where it gets very close to $0.968\pm 0.006$ for $N \approx 55 - 65$ and $50 \lesssim A \lesssim 80 $. While the existence of a negative cosmological constant ($\Lambda < 0$) has been suggested by some authors to solve such 'eternal inflation' problem, this non-conventional solution can't be tested for this model as the slow-roll parameter $\epsilon$ is not a function of $\Lambda$. We have examined the validity of both the classical linear and the new non-linear energy conditions. Although the validity of the linear energy conditions is not generally expected for such an inflationary model with its $\rho^2$ term, we found that both the weak and strong energy conditions are satisfied ($\rho+p>0$ , $\rho+3p>0$) while the dominant energy condition is violated ($\rho-p<0$). For the nonlinear energy conditions, we found that both the flux and determinat conditions are satisfied ($\rho^2 + \sum p_i^2 \geq 0$  and $ \rho . \Pi p_i \geq 0$). The sound speed causality condition $0 \leq \frac{dp}{d\rho} \leq 1$ is not satisfied, it's superluminal ($\frac{dp}{d\rho} > 1$) at early-times and negative ($\frac{dp}{d\rho} < 0$) at late-times.\par

There are still many important aspects that can be discussed within the realm of the current model in future works. For example, although the exotic nature of inflation in the present model is competitor with the standard results found by the Planck satellite, the latter degenerates with a "dark fluid" approach (unifying inflation with dark energy \cite{persp1}) that apparently seems to overcome the issue of producing particles throughout inflation \cite{persp2}. Models like that could be responsible for creating particles, for describing the cosmological constant and for enhancing entanglement entropy \cite{persp3,persp4}. The alternative model to the standard $\Lambda$CDM paradigm in \cite{persp1} assumed the existence of a single fluid, composed by matter only, which exhibits a non-vanishing pressure mimicking the cosmological constant effects and pushes the universe up. The negative pressure in this alternative model arises naturally due to the positiveness of the Helmotz free-energy with no need to introduce a dark energy term by hand. The fine-tuning problem is also overcome since the contributions due to the vacuum energy is canceled out through cold dark matter. future works will deal with these topics within the realm of present model.

\section*{ Data Availability Statement:}
This manuscript has no associated data or the data will not be deposited.
\section*{Acknowledgments}
We are so grateful to the reviewer for his many valuable suggestions and comments that significantly
improved the paper. The IUCAA, Pune, India, is acknowledged by the author ( A. Pradhan) for giving the facility through the Visiting Associateship programmes. 

\end{document}